\documentclass[letterpaper,prl,aps,showpacs,floatfix,twocolumn]{revtex4-1}
\usepackage{graphicx}
\usepackage[ansinew]{inputenc}
\usepackage{array}
\usepackage{color}
\usepackage{amsmath}
\usepackage{amsxtra}
\usepackage{amstext}
\usepackage{amssymb}
\usepackage{latexsym}
\usepackage{dsfont}
\usepackage{verbatim}

\begin{document}

\title{Macroscopic tunneling of a membrane in an optomechanical double-well potential}
\author{L. F. Buchmann$^{1}$}
\author{L. Zhang$^{1,2}$}
\author{A. Chiruvelli$^{1}$}
\author{P. Meystre$^{1}$}
\affiliation{$^1$College of Optical Sciences and B2 Institute, University or Arizona, Tucson, Arizona 85721}
\affiliation{$^2$School of physics and Information Technology, Shaanxi Normal University,
Xi'an 710061, China}
\date{\today}

\begin{abstract}
The macroscopic tunneling of an optomechanical membrane is considered. A cavity mode which couples
quadratically to the membranes position can create highly tunable adiabatic double-well potentials, which
together with the high Q-factors of such membranes render the observation of macroscopic tunneling
possible. A suitable, pulsed measurement scheme using a linearly coupled mode of the cavity for the verification of the effect is studied. 
\end{abstract}

\maketitle

Optomechanical systems have seen a recent surge in experimental and theoretical interest, culminating in the cooling of micromechanical oscillators to within a fraction of a phonon of their quantum  ground state \cite{Oconnell2010,Teufel2011,Chan2011,Safavi2011} and the reaching of the strong coupling between cavity field and mechanical element \cite{Groblacher, Sankey,Teufel2011b}. Despite these successes, few experimental demonstrations of their non-classical behavior have been achieved so far. Notable exceptions include Ref.~\cite{Oconnell2010}, which coupled the mechanical oscillator to a Josephson qubit to detect the presence of a single mechanical phonon, and Ref.~\cite{Safavi2011}, which demonstrated the asymmetry between up-converted and down-converted photons of a probe laser field, an unambiguous signature of the asymmetry between phonon absorption and emission. Additional proposals to generate and exploit nonclassical effects in cavity optomechanics include schemes to squeeze a motional quadrature of the oscillator \cite{Braginsky80,Clerk, Girvin0, Hertzberg, Woolley, Junho}, perform quantum state  tomography \cite{Oconnell2010,Swati,Vanner}, or offer alternative ways to engineer non-classical mechanical  states \cite{Peano,Mazzola} including most intriguingly perhaps the realization of Schr{\"o}dinger cat states in truly macroscopic systems \cite{Romero2010a,Romero2010b,Chang2010}.

This paper extends these considerations by exploring the possibility to realize and monitor the quantum tunneling of an optomechanical system operating deep in the quantum regime through a classically forbidden potential barrier. The observation of the tunneling of such a truly macroscopic object has not been achieved yet, although theoretical proposals have been made~\cite{Hakonen2011}. We find that this can be achieved in a ``membrane-in-the-middle'' (MiM) configuration~\cite{Thompson,Jayich} under conditions that are close to being realizable in current state-of-the-art experiments. We propose a detection scheme based on pulsed optomechanics ideas~\cite{Vanner}  that permits to monitor the tunneling dynamics through a series of weak measurements of the membrane position.

Our approach relies on adiabatically raising a potential barrier, whose parameters can be widely tuned, at the location of a mechanical harmonic oscillator. We show that the ground state of the resulting double-well potential can exhibit tunneling rates several orders of magnitude larger than the decoherence rate of the mechanical membrane, and that a weak optomechanical position measurement is enough to monitor the tunneling. Besides tunneling, the proposed scheme allows for the study of the quantum Zeno effect ~\cite{Sudarshan} in a mechanical context and provides a comparatively simple scheme for the preparation and characterization of non-classical mechanical states of interest for quantum metrology and sensing.

\begin{figure}[t]
\includegraphics[width=0.45 \textwidth]{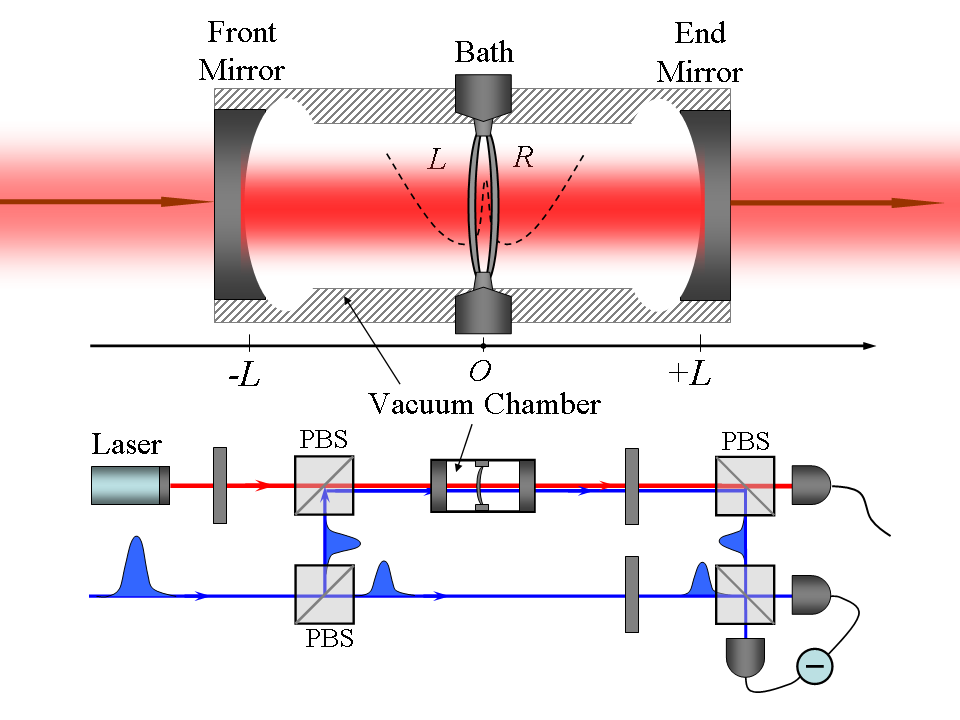}
\caption{(Color Online) Membrane-in-the-middle optomechanical system. The dashed lines illustrate the adiabatic double-well potential induced by the field mode. The membrane in shown in a superposition of left-well state and right-well state. The lower figure sketches a scheme for the detection of membrane tunneling.}
\label{figure1}
\end{figure}

We consider a MiM optomechanical system~\cite{Thompson,Jayich,Pierre1,Pierre2} consisting of two fixed mirrors and a partially reflecting membrane of center-of-mass oscillation frequency $\omega_M$ and effective mass $M$ between them, see Fig.~\ref{figure1}. A unique feature of this system is the option to realize either linear or quadratic optomechanical couplings, depending on the precise membrane's equilibrium position.  Our proposed scheme involves one cavity mode that couples quadratically to the membrane to realize a double well potential, and a second mode that couples linearly and is excited by short optical pulses to first prepare and then monitor the membrane position. The mode frequencies as a function of membrane displacement $x$ have been verified experimentally~\cite{Sankey} and resemble very closely sine and cosine curves~\cite{Thompson,Jayich} for large enough mode numbers.

Consider first the quadratic coupling mode. The corresponding system Hamiltonian is
\begin{equation}
\hat{H}=\hat{H}_{\mathrm{M}}+\hat{H}_{\mathrm{opt}}+\hat{H}_{\mathrm{pump}}+ \hat{H}_{\kappa }+\hat{H}_{\gamma},
\end{equation}
where we have in harmonic oscillator units ($\omega_M = M = 1$)
\begin{eqnarray}
\hat{H}_{\mathrm{M}} &=&\frac{\hat{p}^{2}}{2}+\frac{1}{2}\hat{x}^{2},  \\
\hat{H}_{\mathrm{opt}} &=&\hbar \left( \omega _{c}+\frac{1}{2} g_2 \hat{x}^{2}\right) \hat{a}^{\dagger }\hat{a}, \\
\hat{H}_{\mathrm{pump}} &=&i\hbar \left( \eta e^{-i\omega _{p}t}\hat{a}^{\dagger }-\eta ^{\ast }e^{i\omega _{p}t}\hat{a}\right) .
\end{eqnarray}
Here $\hat{H}_{\mathrm{M}}$ is the energy of the mechanical resonator and $\hat{H}_{\mathrm{opt}}$ is the optomechanical interaction, with $\omega_c$  the resonance frequency of the cavity and $g_2$  the quadratic optomechanical coupling constant. The optical pumping at frequency $\omega_p$ with pumping rate $\eta$ is described by $\hat{H}_{\mathrm{pump}}$. The dissipation terms $\hat{H}_{\kappa }$ and $\hat{H}_{\gamma}$ account  for cavity and mechanical damping with rates $\kappa$ and $\gamma$ respectively.

The Heisenberg equations of motion for the membrane position can be cast as the second order equation
\begin{equation}
\frac{d^{2}\hat{x}}{dt^{2}}+\frac{\gamma}{2}\frac{d\hat{x}}{dt}=
-\hat{x}-g_2\hat{a}^{\dagger }\hat{a} \hat{x}
+\hat{\xi}\left( t\right),
\label{scndorderx}
\end{equation}
where $\hat{\xi}\left( t\right)$ is the noise operator associated with thermal
damping of the oscillator, and the evolution of the cavity field is governed by
\begin{eqnarray}
\frac{d}{dt}\hat{a} =-i\left( \delta _{c}+\frac{1}{2}g_2 \hat{x}^2\right) \hat{a}-\frac{\kappa }{2}\hat{a}  +\eta+\sqrt{\kappa }\hat{a}_{\rm in}.\label{cavitydyn}
\end{eqnarray}
Given stable parameters, the cavity field approaches its steady state in a timescale $\kappa^{-1}$.  For $\kappa \gg ( \omega_M, \gamma)$ it therefore follows adiabatically the position of the membrane (note that this neglects the optomechanical membrane cooling), and the membrane dynamics are robust against fluctuations of the cavity field. Furthermore, a fast cavity decay rate destroys the quantum correlations between the light field and the membrane, in which case it is sufficient to treat the optical field classically, which we do in the following. The intracavity intensity as a function of the membrane position is
\begin{equation}
\bar{I}(\hat{x},t) \approx \frac{\eta ^{2}} {\left(\kappa /2\right)^{2}+\Delta ^{2}\left(\hat{x},t \right) },
\end{equation}
where
$
\Delta \left( \hat{x},t \right) \equiv \delta _{c}+g_{2}\langle\hat{x}^{2}\rangle\left( t \right),
$
with $\delta_c=\omega_c-\omega_p$ the detuning of the pump from the cavity resonance. Inserting this expression into Eq.~(\ref{scndorderx}) and integrating the right hand side gives the effective potential acting on the membrane,
\begin{equation}
U(\hat{x}) =\frac{1}{2}\hat{x}^{2}+\frac{4\eta^{2}}{\kappa }\arctan \left[ \frac{\Delta \left( \hat{x},t \right) }{\kappa
/2}\right].  \label{U}
\end{equation}%
The light field adds an arctangent function to the harmonic potential, which for  positive coupling $g_2$ leads to tighter confinement and for negative couplings, which is the case we are interested in, to a symmetric barrier whose parameters depend on $\eta$ and $\delta_c$.  For $D\equiv-\kappa^2-16\eta^2g_2>0$,
the potential becomes a symmetric double well with two local minima at
\begin{subequations}
\label{distandheight}
\begin{equation}
\pm x_\mathrm{min}=\pm \sqrt{-\frac{2\delta_c+\sqrt{D}}{2g_2}},
\end{equation}
separated by a barrier of height~\cite{footnote}
\begin{equation}
 E_\mathrm{b}=-\frac{1}{2}x_\mathrm{min}^2+\frac{4\eta^2}{\kappa}\left[\arctan\left(2\delta_c/\kappa\right)
+\arctan\left(\sqrt{D}/\kappa\right)\right].
\end{equation}
\end{subequations}

\begin{figure}
\includegraphics[width=7.5cm]{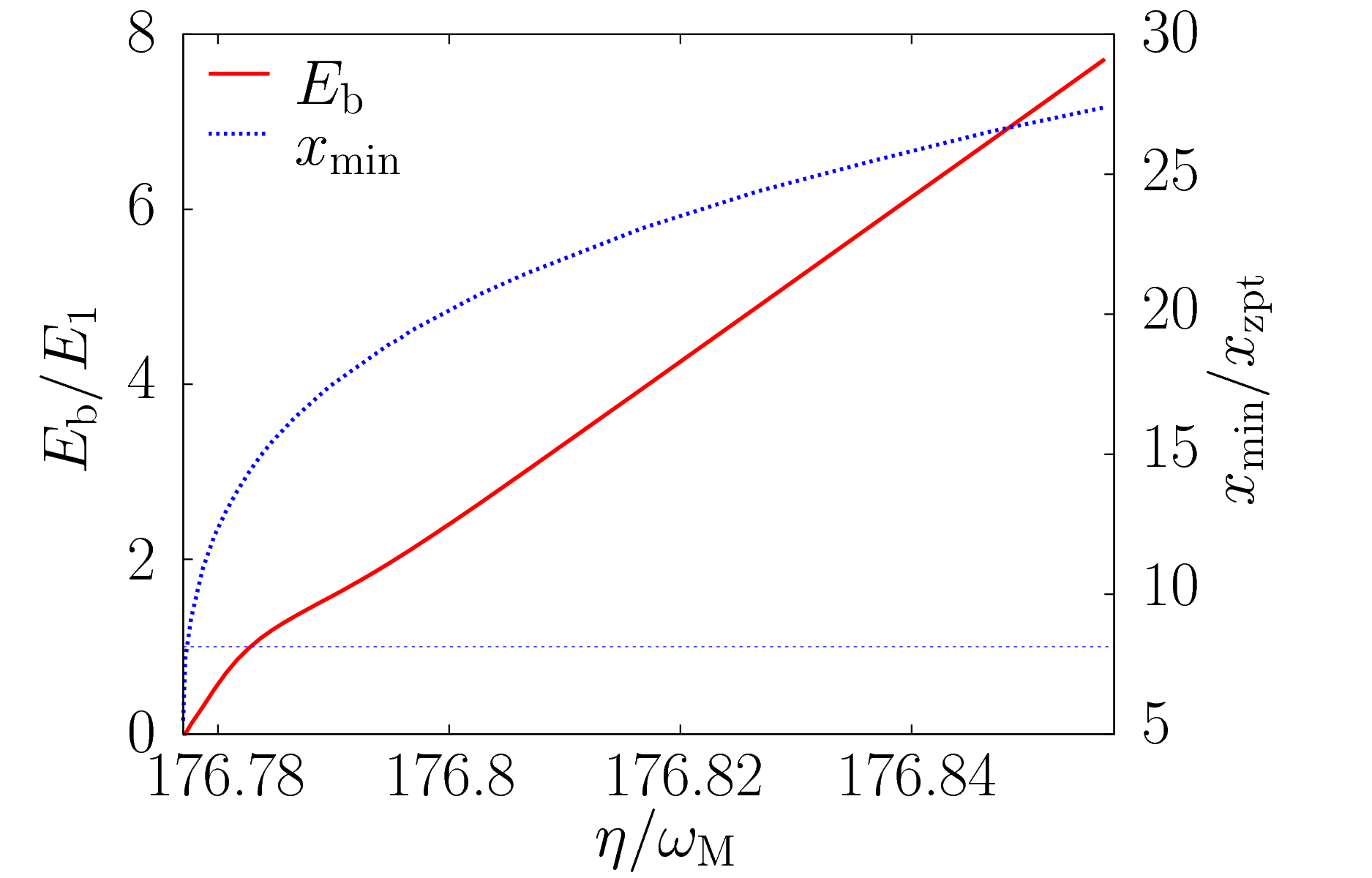}
\caption{(Color Online) Ratio of barrier height to ground state energy (solid line, left axis) and half minimum separation $x_{\rm min}$ in units of the zero-point width $x_{\rm zpt}$ (dotted line, right axis) as functions of the pumping rate $\eta$. Here the coupling strength is $g_2/\omega_\mathrm{M}=-2\times10^{-4}$ and the detuning is zero.}
\label{potparams}.
\end{figure}

Typical parameters allow for a wide range of separations between minima and barrier heights, as illustrated in Fig.~\ref{potparams},  which shows $x_\mathrm{min}$ and $E_{\rm b}$ as a function of the pumping rate $\eta$. The distances between minima can be substantial, and the traps should be accordingly shallow to give reasonable  tunneling rates. For the numerical calculations, we assumed the quadratic single photon coupling $g_2/\omega_M=-2\times 10^{-4}$, which for a typical membrane with $\omega_M/2\pi\approx 100$ kHz translates into a single  photon optomechanical coupling of $20$ Hz.  Such relatively high coupling rates can be obtained by using avoided crossings between higher transverse modes \cite{Sankey}. In addition, a noteworthy aspect of this scheme is its versatility, as weak couplings and/or high decay rates can be compensated by adjusting the detuning or increasing the input power, see Eqs.~(\ref{distandheight}), which for the presented calculations is of the order of 1 $\mu$W. Generally speaking, low effective membrane mass and frequency are desirable to keep the tunneling rates high.

To estimate the tunneling rates in the double-well potential, we consider the membrane Hamiltonian $\mathcal{H}=\sum_{i=1}^\infty E_i\hat{c}_i^\dag\hat{c}_i$,
with eigenenergies $E_i$ determined the time independent Schr\"odinger equation $[ \hat{p}^2/2 + \hat{U}\left(\hat{x}\right) ]\psi=E\psi. $ If $E_1$ and $E_2$ are smaller than the barrier height $E_\mathrm{b}$, their values lie very close together. The corresponding eigenstates $\psi_1$ and $\psi_2$ are symmetric and antisymmetric, respectively, but exhibit similar squared amplitude.
\begin{figure}[h]
\includegraphics[width=0.45 \textwidth]{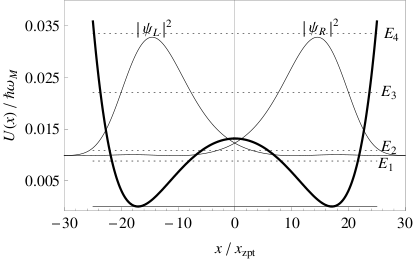}
\caption{Double well and resulting splitting of energy levels. Also shown the squared amplitudes of the two
localized wavefunctions. Parameters used: $g_2/\omega_\mathrm{M}=-2\times 10^{-4}$, $\eta/\omega_\mathrm{M}=176.785$.}
\label{doublewell}
\end{figure}
Thus the states
\begin{equation}
\psi_{R,L}(x)=1/\sqrt{2}(\psi_1(x)\pm \psi_2(x))
\end{equation}
are located predominantly in one of the two potential wells. A typical situation is depicted in Fig.~\ref{doublewell}, which shows a double-well potential, its energy levels as well as the squared amplitudes of the located wave functions $\psi_{L/R}$.

\begin{figure}
\includegraphics[width=7.5cm]{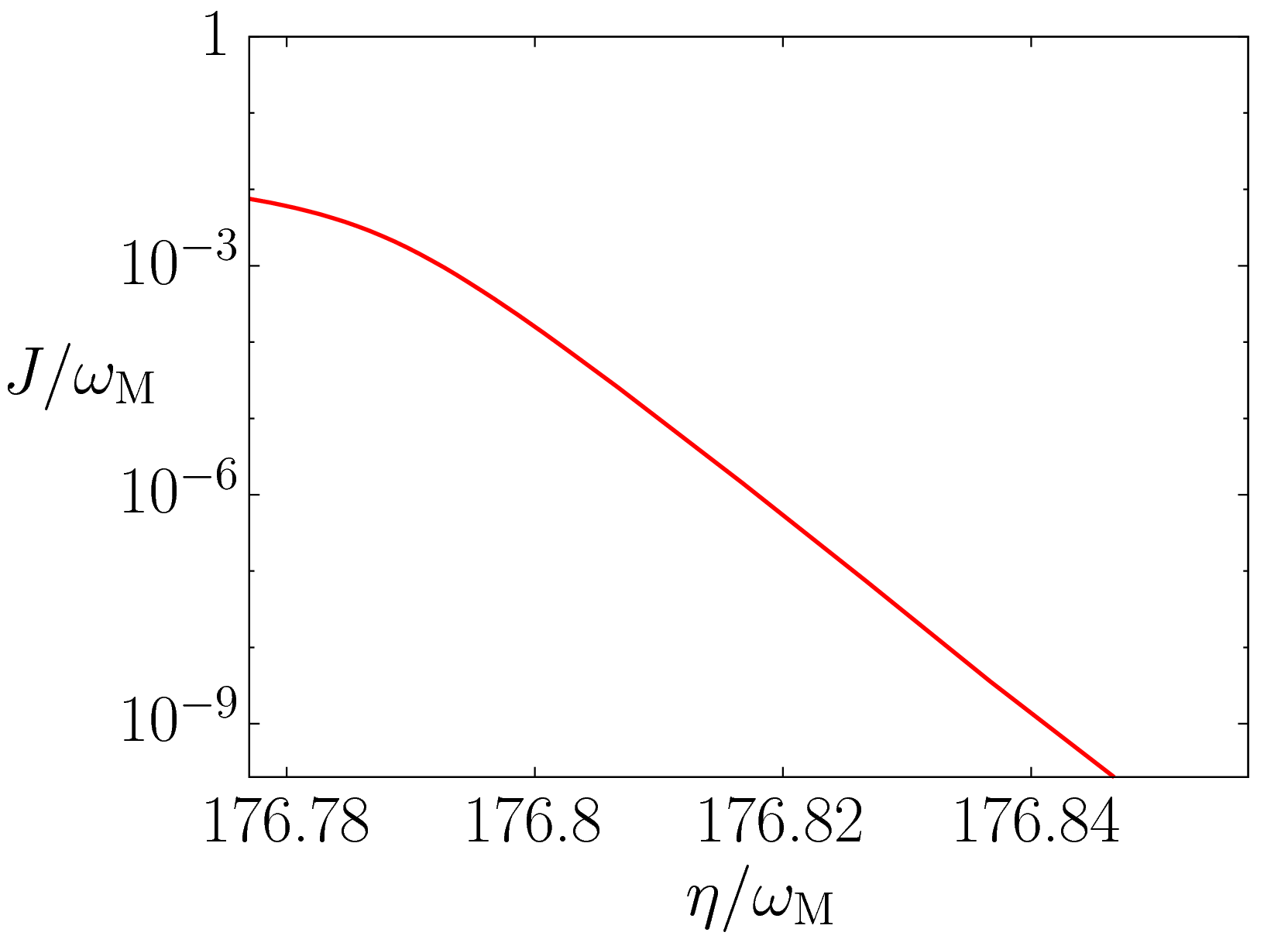}
\caption{(Color Online) Tunneling rate $J$ of the two lowest lying states, normalized to $\omega_M$, as a function of the scaled pumping rate $\eta/\omega_M$ for $g_2/\omega_\mathrm{M}=-2\times10^{-4}$, $\kappa/\omega_\mathrm{M}=10$ and $\delta_c=0$. Note the logarithmic y-axis.}
\label{tunnelrates}
\end{figure}

Reexpressing the membrane Hamiltonian in terms of the localized modes
\begin{equation}
\hat{c}_{R,L}(x)=1/\sqrt{2}(\hat{c}_1\pm \hat{c}_2)
\end{equation}
we obtain
\begin{equation}
\mathcal{H}=\sum_{j\in\{L,R\}}E_j\hat{c}_j^\dag\hat{c}_j+\frac{J}{2}(\hat{c}_L^\dag\hat{c}_R+\hat{c}_R^\dag\hat{c}_L)+\sum_{i>2}E_i\hat{c}_i^\dag\hat{c}_i,
\end{equation}
with $E_L=E_R=(E_1+E_2)/2$ and the tunneling rate $J=E_2-E_1$. Figure~ \ref{tunnelrates} shows the normalized tunneling rate $J/\omega_M$ as a function of the pumping rate $\eta/\omega_M$, illustrating its exponential decrease for increasing $\eta$, i.e., increasing well separation, see Fig.~\ref{potparams}.  If more eigenvalues $E_i$ lie below the barrier they can be split up analogously but we find that ground state tunneling rates become unpractically small in such a situation.

In order for tunneling not to be destroyed by decoherence effects the value of $J$ must be large compared to $\omega_M/Q$. Recent experiments with $\mathrm{Si}_3\mathrm{N}_4$ membranes demonstrated $Q$-factors in excess of $10^6$ \cite{regal} at frequencies of $~500$kHz, with $\omega_\mathrm{M}/Q = 0.5$ sec${}^{-1}$ and effective mass of $\sim 50$ng. For such a membrane,  we find tunneling rates in the range of $J\sim 100$Hz, indicating that the realization of macroscopic optomechanical tunneling should be within reach.

We now turn to a discussion of the detection and monitoring of the tunneling. Continuous and precise monitoring of the membrane's position is possible in principle through a homodyne measurement of the output light from a linearly coupled mode. The effect of such a continuous position measurement has been studied in the context of single atom tunneling \cite{millburn}, and reveals two main effects. Firstly, continuous measurements can prevent the tunneling from happening altogether due to the quantum Zeno effect. The information about the membrane's position continuously projects it towards a position eigenstate and thus suppresses the tunneling. Secondly, the back-action from the measurement can heat up the membrane and  thus eventually supply enough energy to cross the barrier thermally, without tunneling. Although demonstrating the quantum Zeno effect for a macroscopic membrane is certainly noteworthy, we focus here on monitoring the dynamics of the tunneling process. To this end, instead of a continuous measurement we propose to perform a discrete sequence of weak measurements using the pulsed optomechanics technique recently introduced by Vanner {\it et al.}~\cite{Vanner}.

The quadratically coupled resonator mode is not an adequate probe, both because it does not force the membrane into either well due to the ambiguity of the sign of $x$. Instead, we use a second resonator mode that is coupled linearly to the membrane. We assume in the following that the center-of-mass motional mode of the membrane is initially prepared in its ground state, a step that is of course essential and that could be achieved either cryogenically or via an initial optomechanical cooling stage, depending on the specifics of the experimental setup. Following that preparation, a symmetric barrier is  raised adiabatically by turning on the power of the quadratically coupled mode, so that the membrane finds itself in the symmetric  ground state of the double well potential with density operator $\rho=|\psi_1\rangle\langle\psi_1|$. A first measurement pulse then determines the membrane's position through a homodyne measurement, its outcome $x_\mathrm{res}$  projecting the membrane with equal probabilities in the left or right well. That measurement should be just strong enough to project the state of the membrane unambiguously into one of the two localized states to allow for the monitoring of subsequent tunneling. The post-measurement density operator after such a measurement is
\begin{equation}
\rho(t=0)=\frac{\Upsilon\rho\Upsilon^\dag}{\mathrm{Tr}\left(\Upsilon\rho\Upsilon^\dag\right)},
\end{equation}
with
\begin{equation}
\Upsilon=\exp\left[-\frac{(x_\mathrm{res}-\hat{x})^2}{2\sigma^2}\right],
\end{equation}
where $\sigma^2$ is the uncertainty of the measurement, inversely proportional to the measurement  strength \cite{Vanner}. It is determined by the coupling strength of the light mode to the membrane and the pulse intensity.

\begin{figure}
\includegraphics[width=8.5cm]{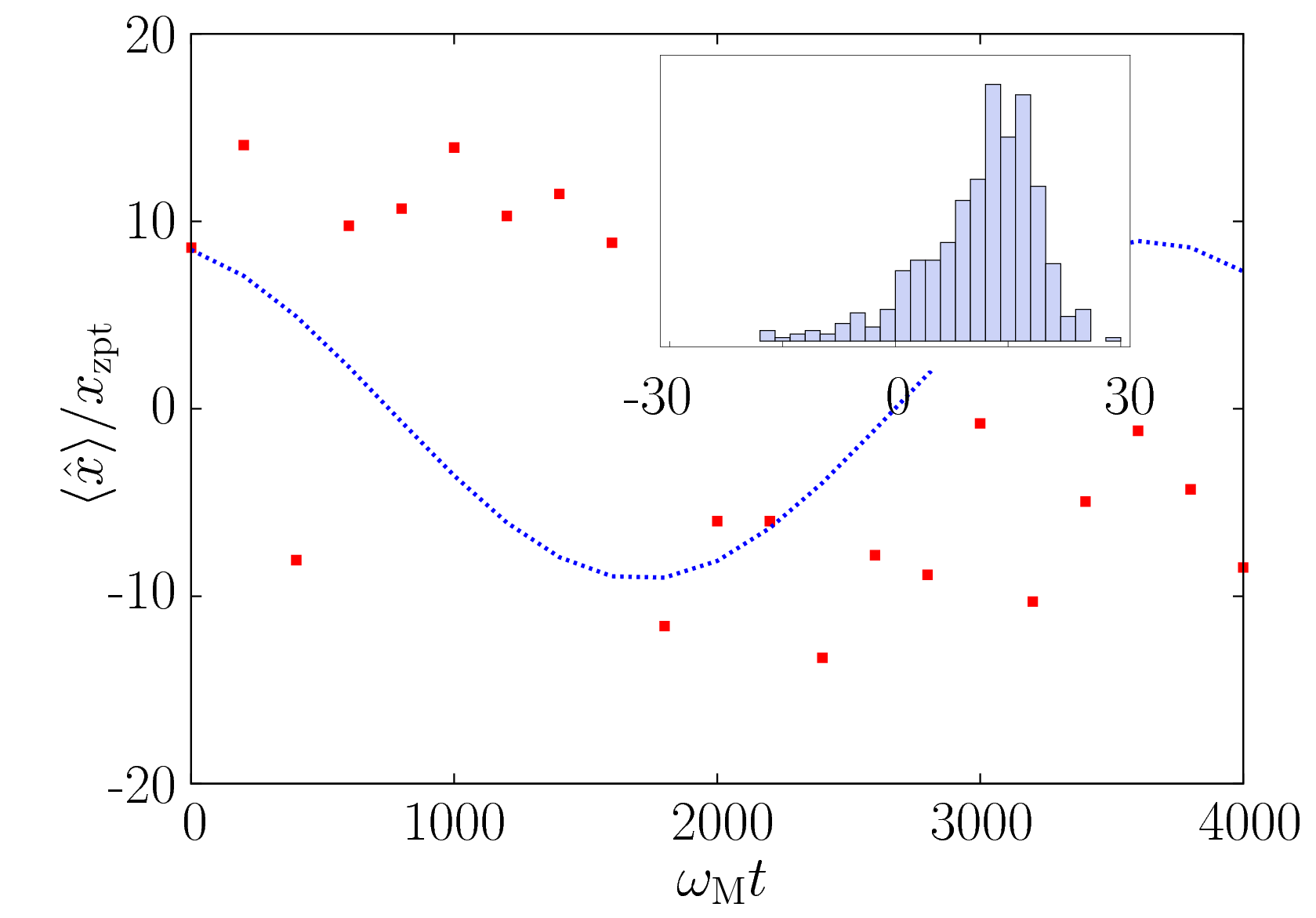}
\caption{(Color Online) Outcomes of a typical sequence of 20 position measurements on the membrane initially in its ground state. The solid line gives the coherent evolution (i.e. without further measurements) of $\langle\hat{x}\rangle$ for this particular trajectory.  The histogram depicts the distribution of 200 measurement outcomes  at $\omega_\mathrm{M}t=20$, where only trajectories which started in the left well were post-selected. Parameters of the potential are the same as in Fig.~\ref{doublewell}, the measurement strength is $1/\sigma=0.02$.}
\label{jumpplot}
\end{figure}

Following that initial preparation, the density operator evolves unitarily, with the membrane starting to tunnel from the random initial well to the opposite one. One way to demonstrate tunneling would be to wait a time of the order of  $J^{-1}$ before the next measurement pulse, assuring the highest probability to find the membrane in the opposite well. To obtain information about the tunneling dynamics, though, it is preferable to perform a sequence of measurements weak enough to keep the membrane's energy below the barrier. The outcome of a numerical simulation of this process for the situation of Fig.~\ref{doublewell} is summarized in Fig.~\ref{jumpplot}. For this simulation, a sequence of 20 measurements were performed in the time $J^{-1}$. The insert shows a histogram of outcomes of the weak measurements of just one time, indicating the spread in values of the membrane position characteristic of weak measurements. We note that throughout the simulation, the energy remained below the barrier height. One can clearly make out the quantum jumps in the position of the membrane. If one were to repeat the measurement sequence a large number of times, post-selecting only those initiated in the same potential well would lead to the emergence of the coherent tunneling dynamics.

In summary, we have investigated the possibility to observe the tunneling of a macroscopic optomechanical membrane in a cavity. The potential barrier is a consequence of the quadratic optomechanical coupling of the membrane to a lossy cavity mode. This set-up is a versatile tool to  create wide, shallow double-well potentials. Using parameters from recent experiments, we find that achievable tunneling rates can exceed the decoherence rates of the membrane by several orders of magnitude. The effect could be verified using weak pulses of linearly coupled light. Besides the possibility of verifying quantum tunneling at unprecedented scales, it would also be a demonstration of the preparation of a non-classical ``Schr\"odinger Cat''  state. Further studies will include investigations of different ways to characterize such mechanical states nondestructively and the potential exploitation of their non-classical nature in  applications such as high-precision quantum metrology, the study of decoherence mechanisms, and the quantum-classical  transition.

\section{acknowledgements}
This work is supported by the US National Science Foundation, the DARPA ORCHID program through a grant from AFOSR, and the US Army Research Office.


\begin{thebibliography}{99}
\bibitem{footnote} If  $\delta_c > -\sqrt{D}/2$, $U(x=0)$ becomes a local minimum and  the potential is a symmetric triple well. Although such a case is interesting in its own right, we focus on the situation of a double-well in this paper.
\end{thebibliography}
\end{document}